\documentclass[journal]{IEEEtran}

\usepackage[cmex10]{amsmath}
\usepackage{amssymb}

\usepackage{cases}
\usepackage{multirow}
\usepackage{empheq}

\usepackage{units}
\usepackage{cite}
\usepackage{verbatim}

\usepackage{algorithm}
\usepackage{algpseudocode}

\usepackage{siunitx}

\ifCLASSINFOpdf
\usepackage[pdftex]{graphicx}
\else
\usepackage[dvips]{graphicx}
\fi
\usepackage[caption=false,font=footnotesize]{subfig}

\usepackage{color}
\usepackage{multirow}

\newtheorem{theorem}{Theorem}
\newtheorem{lemma}[theorem]{Lemma}

\newtheorem{corollary}[theorem]{Corollary}
\newtheorem{definition}[theorem]{Definition}

\newtheorem{remark}[theorem]{Remark}

\newcommand{\vect}[1]{\mathbf{#1}}

\begin{document}
	
\bstctlcite{IEEEexample:BSTcontrol}
	
	\sloppy
	
	\title{Optimizing the Write Fidelity of MRAMs}
	
	
	\author{\IEEEauthorblockN{Yongjune Kim,
			Yoocharn Jeon, 
			Cyril Guyot, and 
			Yuval Cassuto}
	\thanks{
	}
	\thanks{	
		Y. Kim, Y. Jeon, and C. Guyot are with Western Digital Research, Milpitas, CA 95035 USA (e-mail: \{yongjune.kim, yoocharn.jeon, cyril.guyot\}@wdc.com). Y. Cassuto is with the Viterbi Department of Electrical Engineering, Technion -- Israel Institute of Technology, Haifa, Israel, (e-mail: ycassuto@ee.technion.ac.il)}
	}

	
	\maketitle
	
	\begin{abstract}
		Magnetic random-access memory (MRAM) is a promising memory technology due to its high density, non-volatility, and high endurance. However, achieving high memory fidelity incurs significant write-energy costs, which should be reduced for large-scale deployment of MRAMs. In this paper, we formulate an optimization problem for maximizing the memory fidelity given energy constraints, and propose a biconvex optimization approach to solve it. The basic idea is to allocate non-uniform write pulses depending on the importance of each bit position. The fidelity measure we consider is minimum mean squared error (MSE), for which we propose an iterative \emph{water-filling} algorithm. Although the iterative algorithm does not guarantee global optimality, we can choose a proper starting point that decreases the MSE exponentially and guarantees fast convergence. For an 8-bit accessed word, the proposed algorithm reduces the MSE by a factor of 21. 
	\end{abstract}
	
	\section{Introduction}
	
	Magnetic random access memory (MRAM) is a nonvolatile memory technology that has a potential to combine the speed of static RAM (SRAM) and the density of dynamic RAM (DRAM). Furthermore, MRAM technology is attractive since it provides high endurance and complementary metal-oxide-semiconductor (CMOS) compatibility~\cite{Zhu2008magnetoresistive,Kim2015spin}. 
	
	In spite of its attractive features, one of the main challenges is the high energy consumption to write information \emph{reliably} in the memory element~\cite{Zhu2008magnetoresistive,Kim2015spin,Kim2012write}. In an MRAM device, a memory state ``1'' or ``0'' is determined by the magnetic moment orientation of the memory element~\cite{Zhu2008magnetoresistive}. Switching the magnetic moment orientation requires high write current, which introduces write errors when the energy budget is limited~\cite{Kim2015spin}. In addition, high current injection through the tunneling barriers incurs a severe stress and leads to breakdown, which degrades the endurance of MRAM cells~\cite{Kim2012write,Khvalkovskiy2013basic}. Hence, one of the key directions of MRAM research has been toward providing reliable switching with limited energy cost. At the device level, new materials~\cite{Ikeda2010perpendicular,Meng2006spin} or new switching mechanisms~\cite{Nozaki2010voltage,Wang2012electric} have been explored. Several architectural techniques to reduce write energy can be found in~\cite{Kim2012write,Zhou2009energy,Ranjan2015approximate}. 
	
	However, prior efforts have not considered the differential importance of each bit position in error tolerant applications such as signal processing and machine learning (ML) tasks. In these applications, the impact of bit errors depends on bit position, i.e., most significant bits (MSBs) are more important than least significant bits (LSBs)~\cite{Mittal2016survey,Alioto2017energy}. This differential importance has been leveraged to effectively optimize energy in major memory technologies such as SRAMs~\cite{Frustaci2016approximate,Yang2011unequal,Kim2018generalized,Kim2018sram} and DRAMs~\cite{Cho2014edram,Kim2019optimal}.  
	
	In this paper, we provide a \emph{principled} approach to improving MRAM's write fidelity. In error tolerant applications, the mean squared error (MSE) is a more meaningful fidelity metric than the write failure probability (or bit error rate). We formulate a \emph{biconvex optimization} problem to minimize the MSE for a given write energy constraint. Since the write energy and the MSE depend on the write current and the write pulse duration, we attempt to optimize both parameters by solving the biconvex problem.   
	
	Biconvex problem is an optimization problem where the objective function and the constraint set are biconvex~\cite{Gorski2007biconvex}. A common algorithm for solving biconvex problems is \emph{alternate convex search (ACS)}, which updates each variable by fixing another and solving the corresponding convex problem in an iterative manner~\cite{Wendell1976minimization}. We propose an iterative algorithm based on ACS to optimize the write current and the write pulse duration. In addition, we show that the proposed iterative algorithm converges and the convergence speed can be very fast by choosing a proper starting point. 
	
	In general, ACS cannot guarantee the global optimal solution since biconvex problems may have a large number of local minima~\cite{Gorski2007biconvex}. However, we prove that the proposed iterative algorithm can reduce the MSE exponentially by choosing a proper starting point. Furthermore, we show that this starting point guarantees the fastest convergence. We derive analytic expressions of the optimal solutions for each iteration. Since each iteration of the algorithm corresponds to solving convex problems, we rely on the Karush-Kuhn-Tucker (KKT) conditions to derive the optimal solutions. We also provide water-filling interpretations for each iteration.  
	
	Prior optimization studies on voltage swing of SRAMs~\cite{Kim2018generalized,Kim2018sram} and refresh operations of DRAMs~\cite{Kim2019optimal} are similar in spirit, viz. minimizing the MSE for given resource constraints. However, the MRAM write optimization of this work is \emph{non-convex} whereas the formulated problems in \cite{Kim2018generalized,Kim2019optimal} are convex. Hence, we propose the iterative algorithm and analyze convergence and improvement of the optimized MSE. To the best of our knowledge, our work is the first information-theoretic approach to optimization of write pulse parameters of MRAMs. 
	
	The rest of this paper is organized as follows. Section~\ref{sec:mram} explains the basics of MRAM and the challenges of high write energy consumption. Section~\ref{sec:model} introduces the optimization metrics for MRAM write operations. Section~\ref{sec:optimization} formulates optimization problems and provides the iterative algorithm based on ACS. Section~\ref{sec:analysis} provides theoretical analysis on convergence and MSE reduction. Section~\ref{sec:numerical} gives numerical results and Section~\ref{sec:conclusion} concludes. 
	
	\section{Basic Principles of MRAMs}\label{sec:mram}
	
	\subsection{Basics of MRAMs}

\begin{figure}[t]
		\centering
		\includegraphics[width=0.35\textwidth]{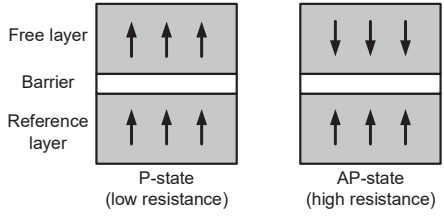}
			\vspace{-4mm}
		\caption{P state and AP state of MTJ MRAM devices.}
		\label{fig:state}
		\vspace{-4mm}
	\end{figure}

	MRAM cells store information by controlling bistable magnetization of ferromagnetic material and retrieve information by sensing resistance of magnetic tunnel junctions (MTJs). An MTJ device consists of two ferromagnetic layers of reference layer (RL) and free layer (FL), separated by a very thin tunneling barrier. RL has a very stable magnetization and it maintains the magnetization throughout all operations, while FL can be switched between two stable magnetization states by a moderate stimulus. The resistance of an MTJ depends on the relative orientation of the FL magnetization with respect to that of the RL as shown in Fig.~\ref{fig:state}. If the magnetizations of FL and RL are in the same direction (parallel- or P-state), then the corresponding resistance is low. The opposite direction (antiparallel- or AP-state) results in high resistance. The difference in tunneling currents between a P-state (low resistance) and a AP-state (high resistance) is utilized to encode binary data~\cite{Zhu2008magnetoresistive,Kim2015spin}.
	
	Writing information into an MTJ is performed by driving a sufficient current through it. Depending on the current's direction, one can flip the magnetization of the FL into P- or AP-state. If a current flows from FL to RL (electrons from RL to FL), electrons are spin-polarized along the magnetization of RL while passing through the layer. The electrons transmitted from the RL interact and exchange the magnetic moments with ones in the FL. If the MTJ is in the AP-state and the current is sufficiently high, then the magnetization orientation is flipped to P-state. When the current is reversed, incoming electrons are polarized along the magnetization of FL. Since the RL's magnetization is parallel to the FL, the majority of the electrons tunnel the barrier while the minority that have antiparallel magnetizations are reflected. Because of this selective tunneling, the antiparallel spins are accumulated in the FL. If the enriched antiparallel spin dominates the FL, it flips the magnetization of the FL into the AP-state.

	The magnetization switching between P state and AP state is not deterministic. The write (switching) failure probability depends on the magnitude and the duration of the write current pulse as follows~\cite[Eq. (26)]{Khvalkovskiy2013basic}: 
	\begin{equation}\label{eq:wfp_complicated}
	p(i,t) = 1 - \exp\left(-\frac{\Delta \pi^2 (i-1)}{4\left\{i \exp(2(i-1)t) - 1 \right\}}\right),
	\end{equation}
	where $\Delta$ denotes the thermal stability factor. The normalized current $i$ is given by $i = \frac{I}{I_{c}}$ where $I$ denotes the actual write current and $I_{c}$ is the critical current. The normalized duration is given by $t = \frac{T}{T_c}$ where $T$ denotes the actual write duration and $T_c$ is the characteristic relaxation time. Note that $\Delta$, $I_{c}$, and $T_c$ are fabrication parameters~\cite{Khvalkovskiy2013basic,Butler2012switching}.
	
	To ensure a low write failure probability, we should control the write current magnitude or the duration judiciously. A longer write duration may lower the write failure probability at the expense of longer write latency and higher energy consumption. Instead of increasing the write duration, we can adopt higher write current. However, it increases the write energy and the risk of dielectric breakdown of the MTJ.
	
	\subsection{Subarray Architecture}
	
	The MRAM cells are arranged in arrays and each of the cells is selectively connected to the read/write circuits to access the data. The metal-oxide-semiconductor field-effect transistors (MOSFETs) are commonly used for the selectors in DRAMs where the required current for memory operations is low enough; a MOSFET with minimum feature sizes can drive the required current. However, the required MRAM write current is more than an order of magnitude higher than that of DRAMs, which requires MOSFETs with large channel width to drive high write current. They are not suitable for high-density memories because of large area on a silicon substrate.

	\begin{figure}[t]
		\centering
		\includegraphics[width=0.46\textwidth]{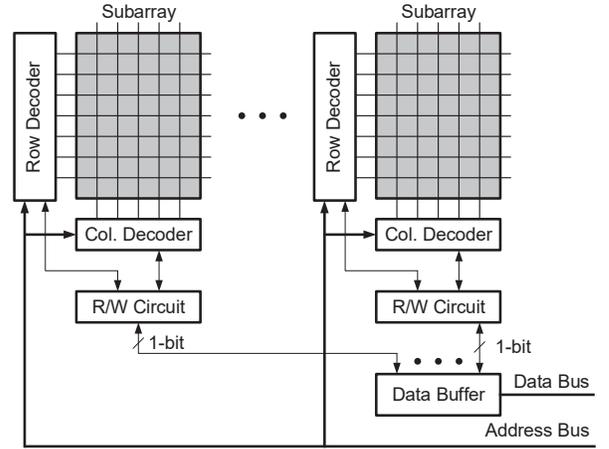}
		\vspace{-4mm}
		\caption{MRAM subarray architecture where each subarray consists of $n_{\text{row}}$ rows and $n_{\text{col}}$ columns.}
		\label{fig:arch}
		\vspace{-4mm}
	\end{figure}
	
	In order to handle this problem, each MRAM cell consists of an MTJ and a threshold switching selector~\cite{Kim2015spin,Yang2017threshold}. These MRAM cells are populated in a crossbar array. To access an MRAM cell, a voltage higher than the threshold voltage of the selector is applied, which turns on the corresponding selector between the selected row-line and column-line, while all the unselected row-lines and column-lines are biased to a midpoint voltage, which keeps all the unselected cells in the array under biases below their threshold voltages. In this manner, the number of needed MOSFETs driving high currents can be reduced from $n_{\text{row}} \times n_{\text{col}}$ to $n_{\text{row}} + n_{\text{col}}$ for a subarray, which is much better suited for high density memories. 
	
	Because of the limited current drivability of the row line and the column line drivers, \emph{only one cell can be accessed at a time in each subarray} unlike DRAMs where a whole page (row-line) can be read/written together (see Fig.~\ref{fig:arch}). Multiple subarrays are operated in parallel to match the required data bandwidth. This MRAM architecture provides an opportunity to write each bit in different conditions (e.g., write current and pulse duration).	
		
	\section{Metrics for MRAM Write Operations}\label{sec:model}

\begin{figure}[t]
		\centering
		\includegraphics[width=0.45\textwidth]{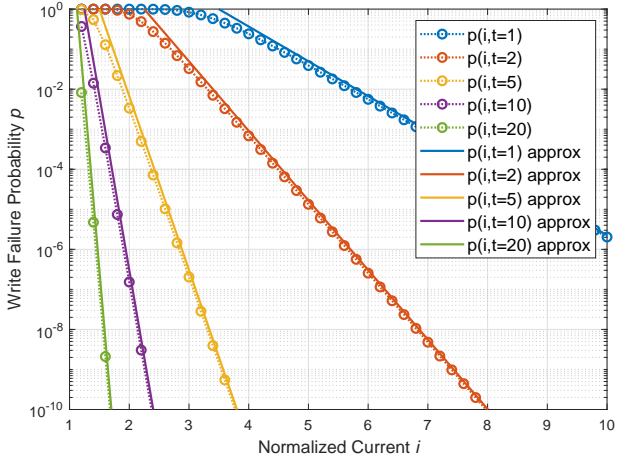}
		\caption{Comparison of the write failure probability \eqref{eq:wfp_complicated} and its approximation \eqref{eq:wfp} ($\Delta = 60$ as in \cite[Fig. 13]{Khvalkovskiy2013basic}).}
		\label{fig:p_approx}
		\vspace{-4mm}
	\end{figure}

	The write failure probability expression of \eqref{eq:wfp_complicated} is too complicated to formulate an optimization problem. Fortunately, we can use the following approximation instead of \eqref{eq:wfp_complicated}:
	\begin{align} 
	p(i,t) & \approx c \exp\left(-2(i-1) t\right). \label{eq:wfp}
	\end{align}
	where $c = \frac{\Delta \pi^2}{4}$. This is a slightly modified approximation of~\cite[Eq. (27)]{Khvalkovskiy2013basic} so as to formulate a biconvex optimization problem. Fig.~\ref{fig:p_approx} shows that the approximated write failure probability \eqref{eq:wfp} is very close to \eqref{eq:wfp_complicated}, especially for lower $p$. The write failure probability can be controlled by the normalized current $i$ and the normalized write duration $t$. 
	
	The normalized energy for writing a single bit is given by
	\begin{equation} \label{eq:energy_bit}
	\mathsf{E}(i,t) = i^2 t. 
	\end{equation} 
	
	As shown in \eqref{eq:wfp} and \eqref{eq:energy_bit}, the write current $i$ and the write duration $t$ are key knobs to control the trade-off between write failure probability and the write energy. If we allocate different write currents and durations depending on the importance of each bit position, then the corresponding current and duration assignments are given by
	\begin{align}
	\vect{i} = (i_0, \ldots, i_{B-1}),\quad 	\vect{t} &= (t_0, \ldots, t_{B-1})
	\end{align}
	where $i_0$ and $t_0$ define the write pulse for least significant bit (LSB) and $i_{B-1}$ and $t_{B-1}$ are the write pulse parameters for most significant bit (MSB).


	We define metrics for energy, latency, and fidelity for writing a $B$-bit word. 
	
	\begin{definition}[Normalized Energy] \label{def:energy}
		The normalized energy of writing a $B$-bit word is given by
		\begin{equation}
		\mathsf{E}(\vect{i},\vect{t}) = \sum_{b=0}^{B-1}{i_b^2 t_b}. 
		\end{equation}
	\end{definition}

	\begin{definition}[Normalized Latency] \label{def:latency}
		The normalized latency of writing a $B$-bit word depends on the maximum write duration among $\vect{t}=(t_0, \ldots, t_{B-1})$, i.e., 
		\begin{equation}
		\mathsf{L}(\vect{t}) = \max \{t_0, \ldots, t_{B-1}\}. 
		\end{equation}		
	\end{definition}

	Note that $\mathsf{E}(\vect{i},\vect{t})$ and $\mathsf{L}(\vect{t})$ are resource metrics. As a fidelity metric, we consider mean squared error (MSE). 
	
	\begin{definition} \label{def:mse}
		The MSE of $B$-bit words is given by
		\begin{align}
		\mathsf{MSE}(\vect{i},\vect{t}) 
		&= \sum_{b=0}^{B-1}{4^b p(i_b, t_b)} \nonumber \\
		&= c \cdot \sum_{b=0}^{B-1}{4^b \exp\left(-2(i_b-1) t_b\right)} \label{eq:mse}
		\end{align} 
		where the weight $4^b$ represents the differential importance of each bit position~\cite{Yang2011unequal,Kim2018generalized}. 
	\end{definition}

	\begin{table}[!t]
		\renewcommand{\arraystretch}{1.5}
		\caption{Resource and Fidelity Metrics for Write Operation}
		\vspace{-2mm}
		\label{tab:comparison}
		\centering
		\begin{tabular}{|c|c|c|}	\hline
			 & Metrics & Remarks \\ \hline \hline
			Energy      &  $\mathsf{E}(\vect{i},\vect{t}) = \sum_{b=0}^{B-1}{i_b^2 t_b}$ & Definition~\ref{def:energy} \\ \hline
			Latency     & $\mathsf{L}(\vect{t}) = \max \{t_0, \ldots, t_{B-1}\}$  & Definition~\ref{def:latency} \\ \hline
			Fidelity    & $\mathsf{MSE}(\vect{i},\vect{t}) = \sum_{b=0}^{B-1}{4^b p(i_b, t_b)} $ & Definition~\ref{def:mse} \\ \hline			
		\end{tabular}
	\end{table} 
	
	Table~\ref{tab:comparison} summarizes the defined metrics for writing a $B$-bit word.   
	
	\section{Optimizing Parameters of Write Operations}\label{sec:optimization}
	
	In this section, we investigate optimization of write operation parameters. First, the optimized current and duration for a single bit will be discussed and then we provide biconvex optimization problems for a $B$-bit word. 
	
	\subsection{Optimized Parameters for Single Bit Write}
	
	First, we note that the normalized current should be greater than 1 for a successful write in \eqref{eq:wfp}. It shows that the write current should be greater than the critical current (i.e., $I > I_c$) so as to switch the direction of magnetization~\cite{Khvalkovskiy2013basic,Butler2012switching}. Then, we can formulate the following optimization problem for single-bit (also multi-bit uniform) write: 
	\begin{equation}
	\begin{aligned} \label{eq:min_wfp}
	& \underset{i, t}{\text{minimize}}
	& & p(i, t) = c\exp\left(-2(i-1) t \right)  \\
	&{\text{subject~to}} & & i^2 t \le \mathcal{E}, \quad i \ge 1 + \epsilon, \quad t \ge 0,
	\end{aligned}
	\end{equation} 
	where $\mathcal{E}$ is a constant corresponding to the given write energy budget. We introduce $\epsilon > 0$ to guarantee $i > 1$. 
	This optimization problem is equivalent to  
	\begin{equation}
	\begin{aligned} \label{eq:min_wfp_qe}
	& \underset{i, t}{\text{maximize}}
	& & (i-1) t   \\
	&{\text{subject~to}} & & i^2 t \le \mathcal{E}, \quad i \ge 1 + \epsilon, \quad t \ge 0.
	\end{aligned}
	\end{equation}
	Note that the objective function $(i-1)t$ is not concave. However, we can readily obtain the optimal $i^*$ and $t^*$ as follows.    
	
	\begin{lemma} \label{thm:min_wfp}
		The optimized current and duration for single bit write are $i^* = 2$ and $t^* = \frac{\mathcal{E}}{4}$, respectively. The corresponding write failure probability is given by
		\begin{equation}\label{eq:p_opt}
		p(i^*, t^*) = c \exp\left(- \frac{\mathcal{E}}{2}\right). 
		\end{equation} 
	\end{lemma}
	\begin{IEEEproof}
		The proof is given in Appendix~\ref{pf:min_wfp}.
	\end{IEEEproof}
	Note that the write failure probability is an exponentially decaying function of $\mathcal{E}$.  
		
	\subsection{Optimized Parameters for $B$-bit Word Writes}
			
	We formulate an optimization problem to determine the currents and durations. For a given write energy constraint, we seek to minimize MSE as follows.  
	
	\begin{equation}
	\begin{aligned} \label{eq:min_mse}
	& \underset{\vect{i}, \vect{t}}{\text{minimize}}
	& & \sum_{b=0}^{B-1}{4^b \exp(-2(i_b - 1)t_b)}  \\
	&{\text{subject~to}} & &  \sum_{b=0}^{B-1}{i_b^2 t_b} \le \mathcal{E} \\
	& & & i_b \ge 1 + \epsilon, \quad t_b \ge 0, \quad b=0,\ldots,B-1
	\end{aligned}
	\end{equation}
	We may include additional constraints such as $\mathsf{L}(\vect{t}) \le \delta$ to guarantee a required write speed performance. Note that $\mathsf{L}(\vect{t}) \le \delta$ is a convex constraint. 
	
	Although the optimization problem \eqref{eq:min_mse} is not convex, we show that \eqref{eq:min_mse} is a \emph{biconvex} optimization problem. Hence, we can find suboptimal solutions via effective algorithms such as \emph{alternate convex search (ACS)}~\cite{Gorski2007biconvex}. 
  
	\begin{definition}[Biconvex Set~\cite{Gorski2007biconvex}] Let $S \subseteq X \times Y$ where $X \subseteq \mathbb{R}^n $ and $Y \subseteq \mathbb{R}^m$ denote two non-empty and convex sets. The set $S$ is defined as a \emph{biconvex set} on $X \times Y$,  if for every fixed $\vect{x} \in X$, $S_\vect{x} \triangleq \left\{\vect{y} \in Y \mid (\vect{x},\vect{y}) \in S \right\}$ is a convex set in $Y$ and for every fixed $\vect{y} \in Y$, $S_\vect{y} \triangleq \left\{\vect{x} \in X \mid (\vect{x},\vect{y}) \in S \right\}$ is a convex set in $X$.
	\end{definition}
	
	\begin{definition}[Biconvex Function~\cite{Gorski2007biconvex}] A function $f:S \rightarrow \mathbb{R}$ is defined as a \emph{biconvex function} on $S$, if for every fixed $\vect{x} \in X$, $f_\vect{x}(\cdot)=f(\vect{x},\cdot):S_\vect{x} \rightarrow \mathbb{R}$ is a convex function on $S_\vect{x}$, and for every fixed $\vect{y} \in Y$, $f_\vect{y}(\cdot)=f(\cdot, \vect{y}):S_\vect{y} \rightarrow \mathbb{R}$ is a convex function on $S_\vect{y}$. 
	\end{definition}
		
	\begin{definition}[Biconvex Problem~\cite{Gorski2007biconvex}] An optimization problem of the following form:
		\begin{equation} \label{eq:biconvex}
		\text{minimize} \left\{f(\vect{x},\vect{y}) \mid (\vect{x},\vect{y})\in S \right\}
		\end{equation}
	is defined as a \emph{biconvex problem}, if the feasible set $S$ is biconvex on $X \times Y$ and the objective function $f$ is biconvex on $S$. 		 
	\end{definition}
	
	\begin{theorem}\label{thm:biconvex}
		The optimization problem \eqref{eq:min_mse} is biconvex. 
	\end{theorem}
	\begin{IEEEproof}
		First, we show that $\sum_{b=0}^{B-1}{i_b^2 t_b} \le \mathcal{E}$ is a biconvex set. Note that $i_b^2 t_b$ is a convex function of $i_b$ for every fixed $t_b \ge 0$. In addition, $i_b^2 t_b$ is a convex function for every fixed $i_b \ge 1 + \epsilon$. Hence, $\sum_{b=0}^{B-1}{i_b^2 t_b} \le \mathcal{E}$ is a biconvex set. 
		
		It is clear that $\exp(-2(i_b - 1)t_b)$ is a biconvex function of $i_b$ and $t_b$. Since the positive weight $4^b$ preserves convexity, the objective function is biconvex. 
	\end{IEEEproof}

	Since \eqref{eq:min_mse} is a biconvex problem, ACS can effectively find a suboptimal solution~\cite{Wendell1976minimization,Gorski2007biconvex}. It alternatively updates variables by fixing one of them and solving the corresponding convex optimization problem. We propose Algorithm~\ref{algo:acs} to optimize the write current $\vect{i}$ and the write duration $\vect{t}$ of the biconvex optimization problem \eqref{eq:min_mse} by using ACS. 
	
	\begin{algorithm}
		\caption{ACS algorithm to solve \eqref{eq:min_mse}} \label{algo:acs}
		\begin{algorithmic}[1]
			\State Choose a starting point $\vect{i}^{(0)}$ from the feasible set $S$ and set $k = 0$. 
			\State For fixed $\vect{i}^{(k)}$, find $\vect{t}^{(k+1)}$ by solving the following convex problem:
			\begin{equation}
			\begin{aligned} \label{eq:min_mse_t}
			& \underset{\vect{t}}{\text{minimize}}
			& & \sum_{b=0}^{B-1}{4^b \exp\left(-2\left(i_b^{(k)} - 1\right)t_b\right)}  \\
			&{\text{subject~to}} & &  \sum_{b=0}^{B-1}{(i_b^{(k)})^2 t_b} \le \mathcal{E} \\
			& & & t_b \ge 0, \quad b=0,\ldots,B-1
			\end{aligned}
			\end{equation}
			\State For fixed $\vect{t}^{(k+1)}$, find $\vect{i}^{(k+1)}$ by solving the following convex problem. 
			\begin{equation}
			\begin{aligned} \label{eq:min_mse_i}
			& \underset{\vect{i}}{\text{minimize}}
			& & \sum_{b=0}^{B-1}{4^b \exp\left(-2(i_b - 1) t_b^{(k+1)} \right)}  \\
			&{\text{subject~to}} & &  \sum_{b=0}^{B-1}{i_b^2 t_b^{(k+1)}} \le \mathcal{E} \\
			& & & i_b \ge 1 + \epsilon, \quad b=0,\ldots,B-1
			\end{aligned}
			\end{equation}
			\State If the point $(\vect{i}^{(k+1)}, \vect{t}^{(k+1)})$ satisfies a stopping criterion, then stop. Otherwise, set $k:=k+1$ and go back to line 2. 
		\end{algorithmic}
	\end{algorithm}

	\begin{remark}[Starting Point]
		Since biconvex optimization problems may have a large number of local minima~\cite{Gorski2007biconvex}, a starting point $\vect{i}^{(0)}$ can affect the final solution. We can choose $\vect{i}^{(0)} = (2, \ldots, 2)$ as a starting point, which minimizes the uniform write failure probability (see Lemma~\ref{thm:min_wfp}). In Corollary~\ref{thm:convergence_fast}, we show that this starting point guarantees the fastest convergence.  		 
	\end{remark}

	\begin{remark}[Stopping Criterion~\cite{Gorski2007biconvex}]
		There are several ways to define the stopping criterion in Algorithm~\ref{algo:acs}. For example, we can consider the absolute values of the differences between $(\vect{i}^{(k)}, \vect{t}^{(k)})$ and $(\vect{i}^{(k+1)}, \vect{t}^{(k+1)})$ or the difference between $\mathsf{MSE}(\vect{i}^{(k)}, \vect{t}^{(k)})$ and $\mathsf{MSE}(\vect{i}^{(k+1)}, \vect{t}^{(k+1)})$. Alternatively, we can set a maximum number of iterations. 
	\end{remark}

	\section{Analysis of Alternate Convex Search for MRAM Write Parameters} \label{sec:analysis}

	\subsection{Optimal Solutions for Each Iteration}
	
	In this subsection, we present the optimal solutions for \eqref{eq:min_mse_t} and \eqref{eq:min_mse_i}. Since these problems are convex, we exploit the structure of the problems to derive the optimal solutions analytically using the KKT conditions. 
	
	\begin{theorem}\label{thm:min_mse_t}
		For fixed $\vect{i}^{(k)} = \vect{i}$, the optimal $\vect{t}^{(k+1)} = \vect{t}^*$ of \eqref{eq:min_mse_t} is given by
		\begin{equation}\label{eq:min_mse_sol_t}
		t_b^* =
		\begin{cases}
		0, & \text{if}\: \nu \ge \frac{2 \cdot 4^b \left(i_b - 1\right)}{i_b^2} ; \\
		\frac{\log{\left(\frac{1}{\nu}\cdot \frac{2 \cdot 4^b \left(i_b - 1\right)}{i_b^2} \right)}}{2 \left(i_b - 1\right)}, & \text{otherwise}
		\end{cases}
		\end{equation}
		where $\nu$ is a dual variable of corresponding KKT conditions. Note that $\nu$ depends on the energy budget $\mathcal{E}$.
	\end{theorem}
	\begin{IEEEproof}
		We define the Lagrangian $L_1(\vect{t}, \nu, \boldsymbol{\lambda})$ associated with problem \eqref{eq:min_mse_t} as
		\begin{align}
		L_1(\vect{t}, \nu, \boldsymbol{\lambda}) & = \sum_{b=0}^{B-1}{4^b \exp(-2(i_b - 1)t_b)} \nonumber \\
		&+ \nu \left( \sum_{b=0}^{B-1}{i_b^2 t_b} - \mathcal{E}\right) - \sum_{b=0}^{B-1}{\lambda_b t_b } 	
		\end{align}
		where $\nu$ and $\boldsymbol{\lambda}=(\lambda_0,\ldots,\lambda_{B-1})$ are the dual variables. The details of the proof are given in Appendix~\ref{pf:min_mse_t}. 
	\end{IEEEproof}

	\begin{theorem}\label{thm:min_mse_i}
		For fixed $\vect{t}^{(k+1)} = \vect{t}$, the optimal $\vect{i}^{(k+1)} = \vect{i}^*$ of \eqref{eq:min_mse_i} is given by
		\begin{equation}\label{eq:min_mse_sol_i}
		i_b^* =
		\begin{cases}
		1 + \epsilon, & \text{if}\: \nu' \ge \frac{4^b}{1 + \epsilon} e^{-2t_b \epsilon}  ; \\
		\frac{1}{2t_b}W\left(\frac{2\cdot 4^b t_b e^{2t_b}}{\nu'}\right), & \text{otherwise}
		\end{cases}
		\end{equation}
		where $\nu'$ is a dual variable. Also, $W(\cdot)$ denotes the \emph{Lambert W function} (i.e., the inverse function of $f(x) = xe^x$)~\cite{Corless1996lambert}. 
	\end{theorem}
	\begin{IEEEproof}
			We define the Lagrangian $L_2(\vect{i}, \nu', \boldsymbol{\lambda}')$ associated with problem \eqref{eq:min_mse_i} as
		\begin{align}
		L_2(\vect{i}, \nu', \boldsymbol{\lambda}') & = \sum_{b=0}^{B-1}{4^b e^{-2(i_b - 1)t_b}} + \nu' \left( \sum_{b=0}^{B-1}{i_b^2 t_b} - \mathcal{E}\right) \nonumber \\
		& - \sum_{b=0}^{B-1}{\lambda_b' \left\{i_b - (1 + \epsilon)\right\} } 	
		\end{align}
		where $\nu'$ and $\boldsymbol{\lambda}'=(\lambda_0',\ldots,\lambda_{B-1}')$ are the dual variables. The details of the proof are given in Appendix~\ref{pf:min_mse_i}.		
	\end{IEEEproof}

{
	\begin{remark}
	The solutions of \eqref{eq:min_mse_sol_t} and \eqref{eq:min_mse_sol_i} can be interpreted as water-filling. Each bit position can be regarded as an individual channel among $B$ parallel channels as in \cite{Kim2018generalized,Kim2018sram}. The ground levels depend on the importance of bit positions; hence larger current or longer duration are assigned to more significant bit positions. 
	\end{remark}
}
	
	\subsection{Convergence of MSE}

	We show that Algorithm~\ref{algo:acs} guarantees convergence to a locally optimal MSE. The converged MSE depends on a starting point. 

	\begin{lemma}\label{thm:monotone}The sequence $\left\{\mathsf{MSE}(\vect{i}^{(k)}, \vect{t}^{(k)})\right\}_{k \in \mathbb{N}}$ obtained by Algorithm~\ref{algo:acs} is monotonically decreasing, i.e., $\mathsf{MSE}(\vect{i}^{(k+1)}, \vect{t}^{(k+1)}) \le \mathsf{MSE}(\vect{i}^{(k)}, \vect{t}^{(k)})$ for all $k \in \mathbb{N}$. 
	\end{lemma}
	\begin{IEEEproof} Note that $\mathsf{MSE}(\vect{i}^{(k)}, \vect{t}^{(k+1)}) \le \mathsf{MSE}(\vect{i}^{(k)}, \vect{t}^{(k)})$ and $\mathsf{MSE}(\vect{i}^{(k+1)}, \vect{t}^{(k+1)}) \le \mathsf{MSE}(\vect{i}^{(k)}, \vect{t}^{(k+1)})$ because of \eqref{eq:min_mse_t} and \eqref{eq:min_mse_i}, respectively. Hence, $\mathsf{MSE}(\vect{i}^{(k+1)}, \vect{t}^{(k+1)}) \le \mathsf{MSE}(\vect{i}^{(k)}, \vect{t}^{(k)})$. 
	\end{IEEEproof}
	
	\begin{theorem}\label{thm:convergence}The sequence $\left\{\mathsf{MSE}(\vect{i}^{(k)}, \vect{t}^{(k)})\right\}_{k \in \mathbb{N}}$ obtained by Algorithm~\ref{algo:acs} converges monotonically. 
	\end{theorem}
	\begin{IEEEproof} It is clear that $\mathsf{MSE}(\vect{i}^{(k)}, \vect{t}^{(k)}) \ge 0$ for all $k \in \mathbb{N}$ by \eqref{eq:wfp} and \eqref{eq:mse}. Then, $\left\{\mathsf{MSE}(\vect{i}^{(k)}, \vect{t}^{(k)})\right\}_{k \in \mathbb{N}}$ is monotonically decreasing and bounded below, $\left\{\mathsf{MSE}(\vect{i}^{(k)}, \vect{t}^{(k)})\right\}_{k \in \mathbb{N}}$ converges because of monotone convergence theorem. 
	\end{IEEEproof}

	\begin{corollary}\label{thm:convergence_fast}
		By setting $\vect{i}^{(0)} = (2, \ldots, 2)$, we obtain
		\begin{equation}
		\lim_{k \rightarrow \infty} \left(\vect{i}^{(k)}, \vect{t}^{(k)} \right) = \left(\vect{i}^{(0)}, \vect{t}^{(1)} \right), 
		\end{equation}
		if $t_b^{(1)} \ne 0$ for all $b \in [0, B-1]$. 
	\end{corollary}	
	\begin{IEEEproof}
		We will show that $(\vect{i}^{(0)}, \vect{t}^{(1)})$ (i.e., the solution of \eqref{eq:min_mse_t}) satisfies the KKT conditions of \eqref{eq:min_mse_i}. Then, $\vect{i}^{(1)} = \vect{i}^{(0)}$, which makes Algorithm 1 converge in one step. The details of the proof are given in Appendix~\ref{pf:convergence_fast}. 
	\end{IEEEproof}

	Corollary~\ref{thm:convergence_fast} means that the starting point  $\vect{i}^{(0)} = (2, \ldots, 2)$ guarantees the \emph{fastest convergence}. Note that we do not need to solve \eqref{eq:min_mse_i}.

	\subsection{Starting Point of $\vect{i}^{(0)} = (2, \ldots, 2)$}
	
	The starting point $\vect{i}^{(0)} = (2, \ldots, 2)$ guarantees the fastest convergence. Note that it minimizes the write failure probability for the single bit case (see Lemma~\ref{thm:min_wfp}). In this subsection, we show that $\vect{i}^{(0)} = (2, \ldots, 2)$ is a good starting point, in the sense that it reduces the MSE exponentially with $B$. 
	
	Suppose that the starting point is $\vect{i}^{(0)} = (2, \ldots, 2)$. By Theorem~\ref{thm:min_mse_t} and Corollary~\ref{thm:convergence_fast}, Algorithm 1 provides the following optimized write durations $\vect{t}^{(1)} = \widetilde{\vect{t}} = (\widetilde{t}_0, \ldots, \widetilde{t}_{B-1})$ where 
	\begin{equation}\label{eq:min_mse_sol_t_i0}
	\widetilde{t}_b =
	\begin{cases}
	0, & \text{if}\: \nu \ge \frac{4^b}{2} ; \\
	\frac{1}{2}\log{\left(\frac{1}{\nu}\cdot \frac{4^b }{2} \right)}, & \text{otherwise}.
	\end{cases}
	\end{equation}
	
	\begin{lemma}\label{thm:positive_t}
		If $\mathcal{E} > 2B(B-1)\log{2}$, then $\widetilde{t}_b > 0$ for all $b \in [0, B-1]$ and 
		\begin{equation} \label{eq:positive_t}
		\widetilde{t}_b = \frac{\mathcal{E}}{4B} + \left(b - \frac{B-1}{2}\right) \cdot \log{2}. 
		\end{equation}  
	\end{lemma}
	\begin{IEEEproof}
	The proof is given in Appendix~\ref{pf:mse_improvement}. 
	\end{IEEEproof}

	\begin{theorem}\label{thm:mse_improvement}
		If $\mathcal{E} > 2B(B-1)\log{2}$, then the MSE reduction ratio by Algorithm~\ref{algo:acs} is given by
		\begin{equation} \label{eq:mse_improvement}
		\gamma = \frac{ \mathsf{MSE}\left(\vect{i}^{(0)}, \widetilde{\vect{t}} \right)} {\mathsf{MSE} \left(\vect{i}^{(0)}, \vect{t}^{(0)} \right) } = \frac{3B}{2} \cdot \frac{ 2^B}{4^{B}- 1} \approx \frac{3B}{2} \cdot 2^{-B}
		\end{equation}
		where $\mathsf{MSE}(\vect{i}^{(0)}, \widetilde{\vect{t}} )$ (i.e., the optimized MSE by Algorithm~\ref{algo:acs}) is given by
		\begin{equation}
		\mathsf{MSE}\left(\vect{i}^{(0)}, \widetilde{\vect{t}} \right) = c \cdot \frac{B}{2} \cdot 2^{B} \exp\left(- \frac{\mathcal{E}}{2B}\right)
		\end{equation}		
		where the optimized $\widetilde{\vect{t}}$ is given by \eqref{eq:min_mse_sol_t_i0}. In addition, $\mathsf{MSE} (\vect{i}^{(0)}, \vect{t}^{(0)} )$ (i.e., the MSE by uniform energy allocation) is given by
		\begin{equation}
		\mathsf{MSE} \left(\vect{i}^{(0)}, \vect{t}^{(0)} \right) = c \cdot \frac{4^B - 1}{3} \exp\left(-\frac{\mathcal{E}}{2B}\right)	\end{equation}		
		where $\vect{t}^{(0)}$ is the uniform value to satisfy the energy constraint (i.e., $\vect{t}^{(0)} = \frac{\mathcal{E}}{4B} \cdot (1, \ldots, 1)$). 
	\end{theorem}
	\begin{IEEEproof}
		The proof is given in Appendix~\ref{pf:mse_improvement}. 
	\end{IEEEproof}
	Note that $\mathsf{MSE} \left(\vect{i}^{(0)}, \vect{t}^{(0)} \right)$ is the MSE corresponding to the parameters minimizing the write failure probability (see Lemma~\ref{thm:min_wfp}).
	
	\begin{remark}
		By setting $\vect{i}^{(0)} = (2, \ldots, 2)$, Algorithm~\ref{algo:acs} reduces the MSE exponentially with $B$, compared to the parameters optimized for write failure probability. Although we cannot guarantee that $(\vect{i}^{(0)}, \vect{t}^{(1)} = \widetilde{\vect{t}})$ is globally optimal, $(\vect{i}^{(0)}, \vect{t}^{(1)})$ decrease the MSE exponentially by solving \eqref{eq:min_mse_t} once (see Corollary~\ref{thm:convergence_fast}). Furthermore, the solution of \eqref{eq:min_mse_t} can be easily computed by Lemma~\ref{thm:positive_t}.   
	\end{remark}

	\section{Numerical Results}\label{sec:numerical}
	
	We evaluate the solutions to optimize the write failure probability for single bits as well as the MSE for $B$-bit words. The critical current $I_c$ and the characteristic relaxation time $T_c$ do not affect the numerical results because the normalized values $i = \tfrac{I}{I_c}$ and $t =\frac{T}{T_c}$ are considered. As in \cite{Khvalkovskiy2013basic}, we set $\Delta = 60$ for the thermal stability factor.  
	
	\begin{figure}[t]
	\centering
	\includegraphics[width=0.45\textwidth]{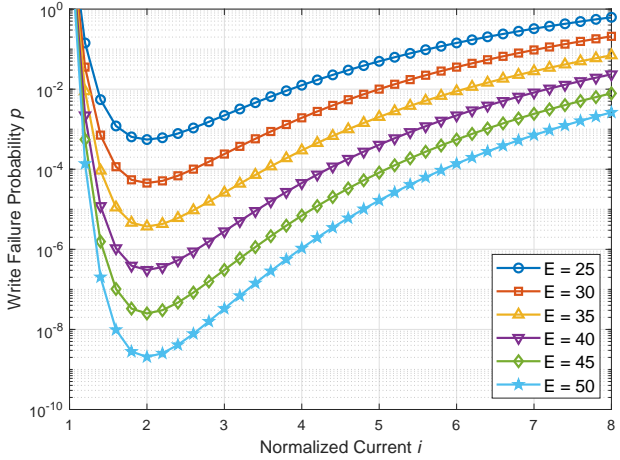}
	\caption{Normalized write current to minimize the write failure probability (see Lemma~\ref{thm:min_wfp}) for several energy constraints.}
	\label{fig:p_opt}
	\end{figure}	

	\begin{figure}[t!]
	\centering
	\subfloat[]{\includegraphics[width=0.45\textwidth]{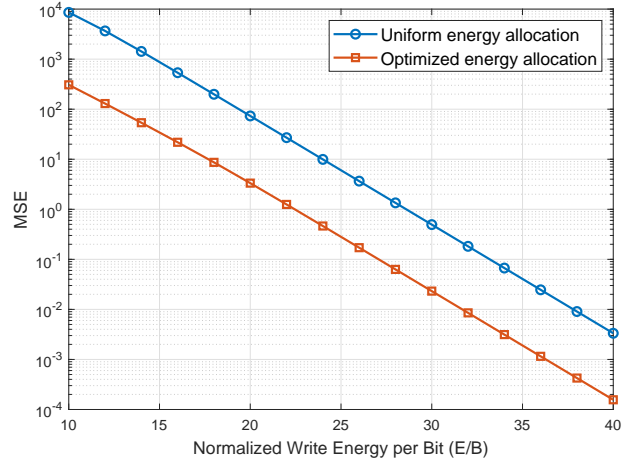}
		\label{fig:optimal_mse}}
	\hfil
	\subfloat[]{\includegraphics[width=0.45\textwidth]{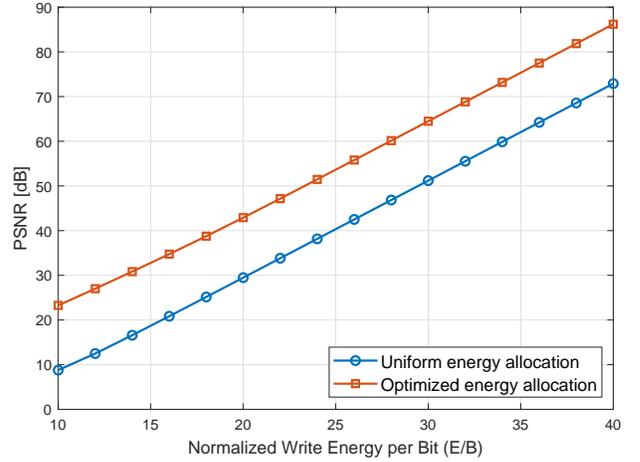}
		\vspace{-4mm}
		\label{fig:optimal_psnr}}
	\caption{Comparison of the conventional uniform energy allocation and the optimized energy allocation by Algorithm~\ref{algo:acs} ($B = 8$): (a) MSE and (b) PNSR.}
	\label{fig:optimal}
	\end{figure}

	Fig.~\ref{fig:p_opt} shows that $i^* = 2$ and $t^* = \frac{\mathcal{E}}{4}$ minimize the write failure probability as proved in Lemma~\ref{thm:min_wfp}. The corresponding minimal write failure probability decreases exponentially with the write energy as shown in \eqref{eq:p_opt}. 
	
	\begin{figure}[t!]
	\centering
	\includegraphics[width=0.45\textwidth]{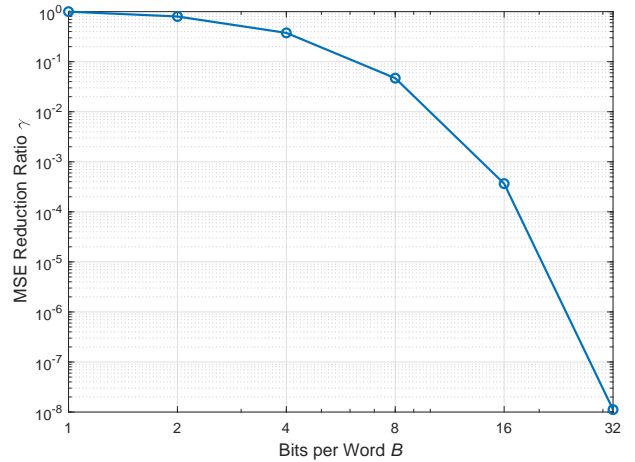}
	\caption{The MSE reduction ratio $\gamma$ by Theorem~\ref{thm:mse_improvement}.}
	\label{fig:mse_improvement}
	\end{figure}

	Fig.~\ref{fig:optimal} shows numerical results by solving \eqref{eq:min_mse}. Fig.~\ref{fig:optimal}\subref{fig:optimal_mse} compares the MSEs of uniform write energy allocation and the optimized energy allocation by Algorithm~\ref{algo:acs}. We set a starting point $\vect{i}^{(0)} = (2, \ldots,2)$. As shown in Theorem~\ref{thm:mse_improvement}, the MSE reduction ratio is $\gamma \approx \frac{3B}{2}\cdot2^{-B} = 0.0469$ for $B = 8$. 	Fig.~\ref{fig:optimal}\subref{fig:optimal_psnr} compares the peak signal-to-noise ratios (PSNRs), which is a widely used fidelity metric for image and video quality. The PSNR depends on the MSE as $\mathsf{PSNR} = 10 \log_{10}{\frac{(2^B-1)^2}{\mathsf{MSE}}}$. At $\mathsf{PSNR} = \SI{40}{dB}$, the optimized write energy allocation can reduce the write energy by \SI{24}{\%}. 

	Fig.~\ref{fig:mse_improvement} shows that the MSE reduction ratio improves exponentially with $B$ (as derived in Theorem~\ref{thm:mse_improvement}). Although we cannot guarantee the optimality, the proposed Algorithm~\ref{algo:acs} is very effective to reduce the MSE. Note that $\gamma = 3.66 \times 10^{-4}$ for $B = 16$ and $\gamma = 1.12 \times 10^{-8}$ for $B=32$.


	\begin{figure*}
	\begin{centering}
		\subfloat[]{\includegraphics[width=0.45\textwidth]{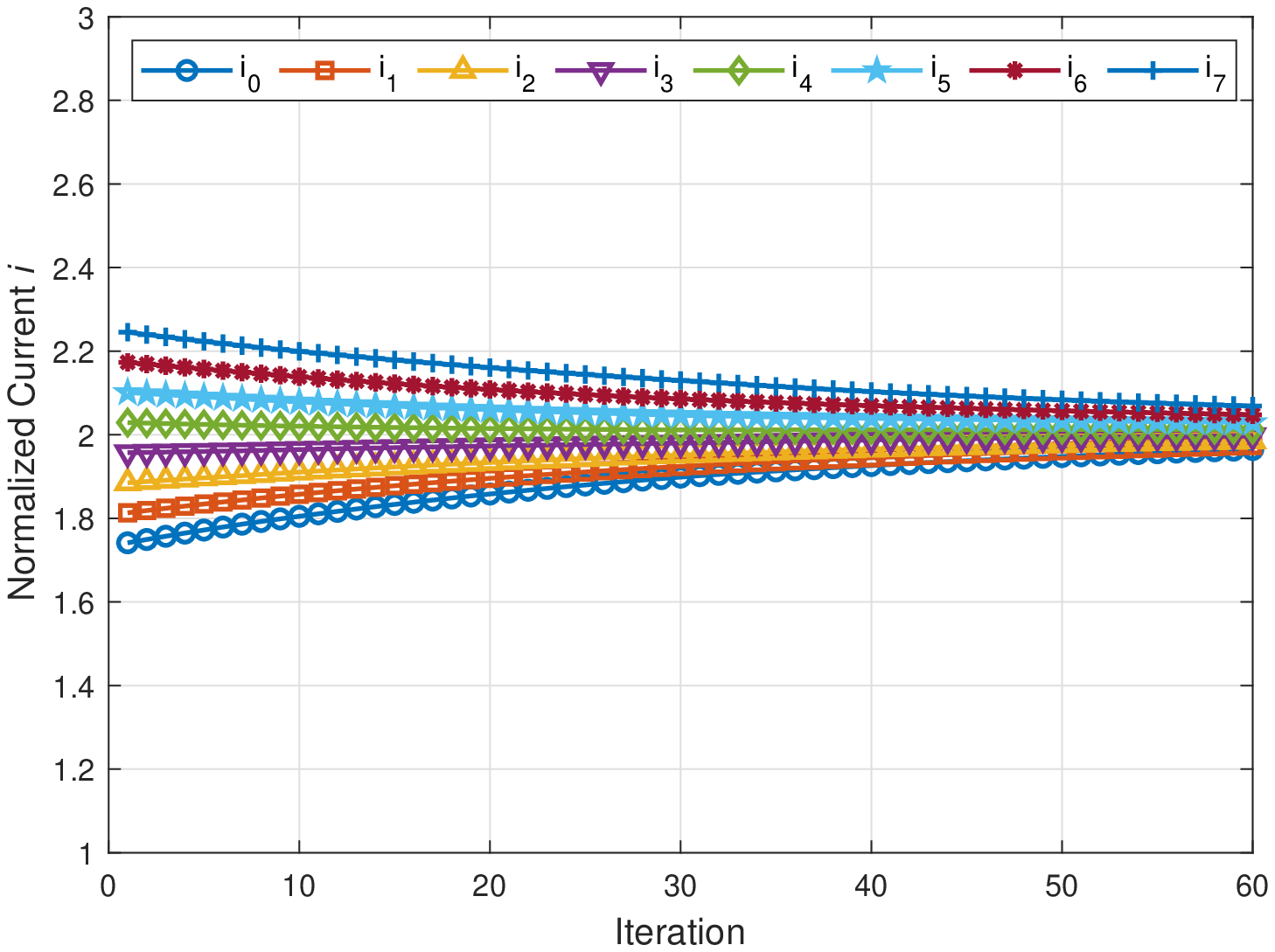}}
		\hfill
		\subfloat[]{\includegraphics[width=0.45\textwidth]{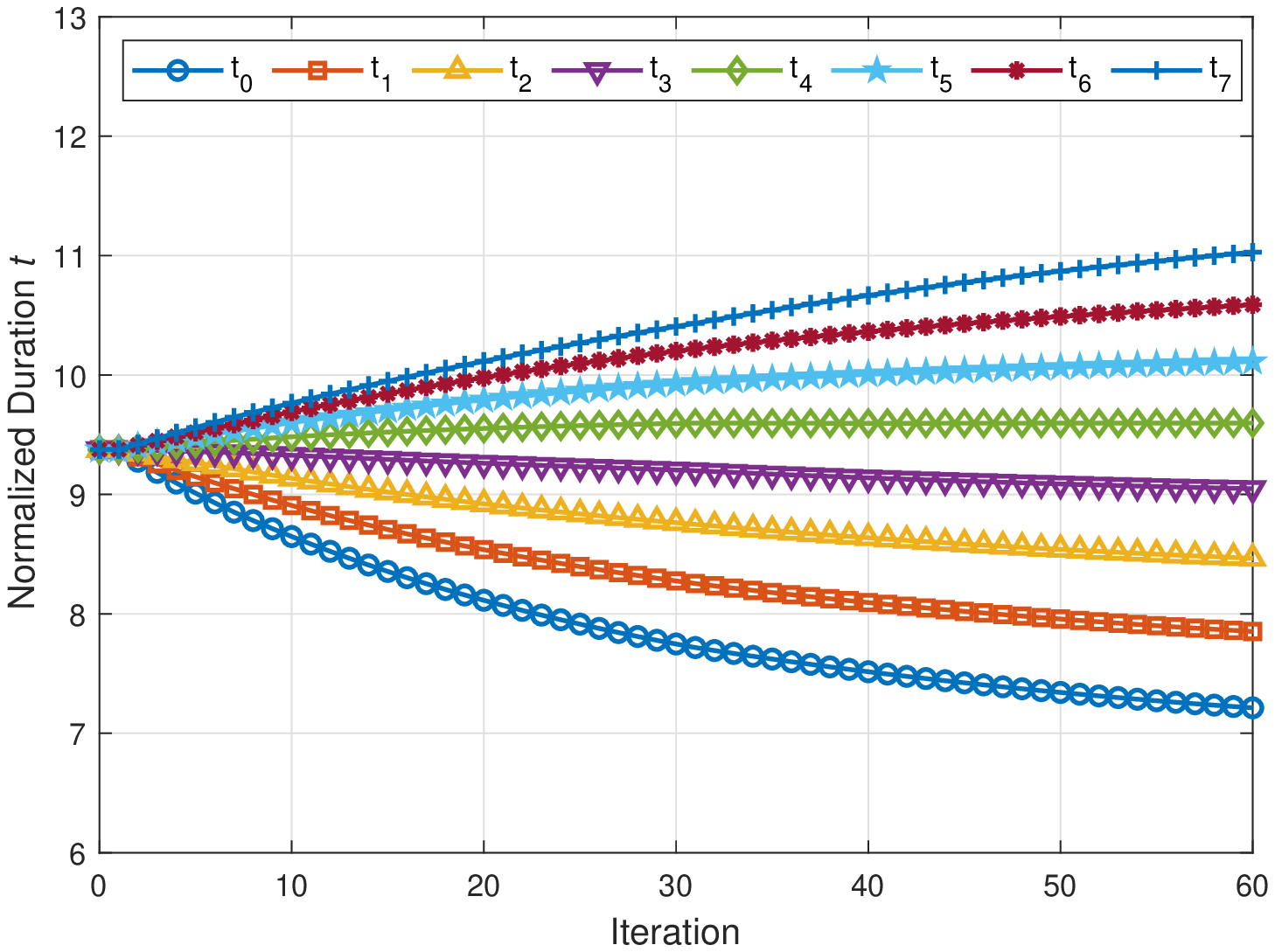}}	

		\subfloat[]{\includegraphics[width=0.45\textwidth]{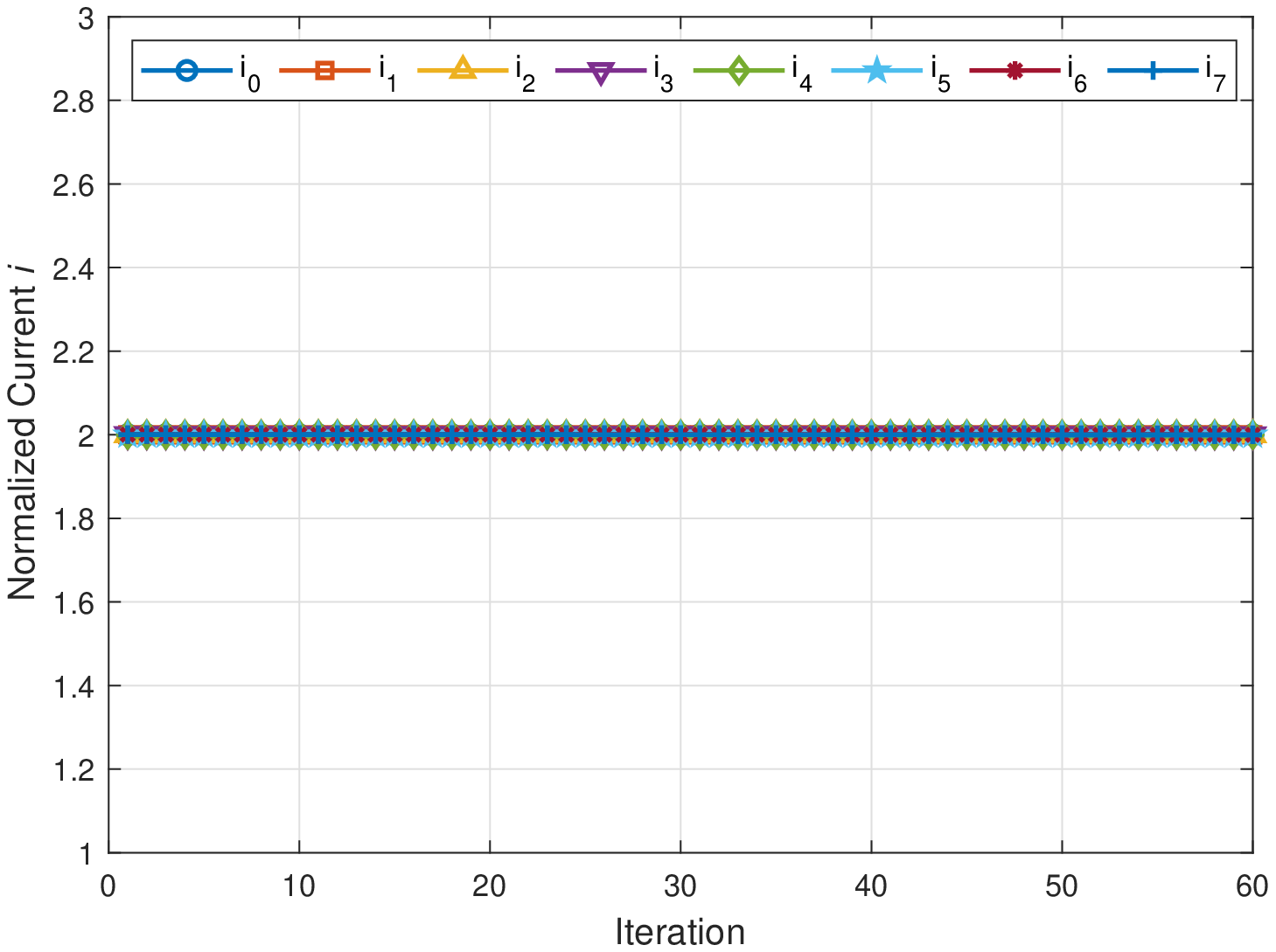}}
		\hfill
		\subfloat[]{\includegraphics[width=0.45\textwidth]{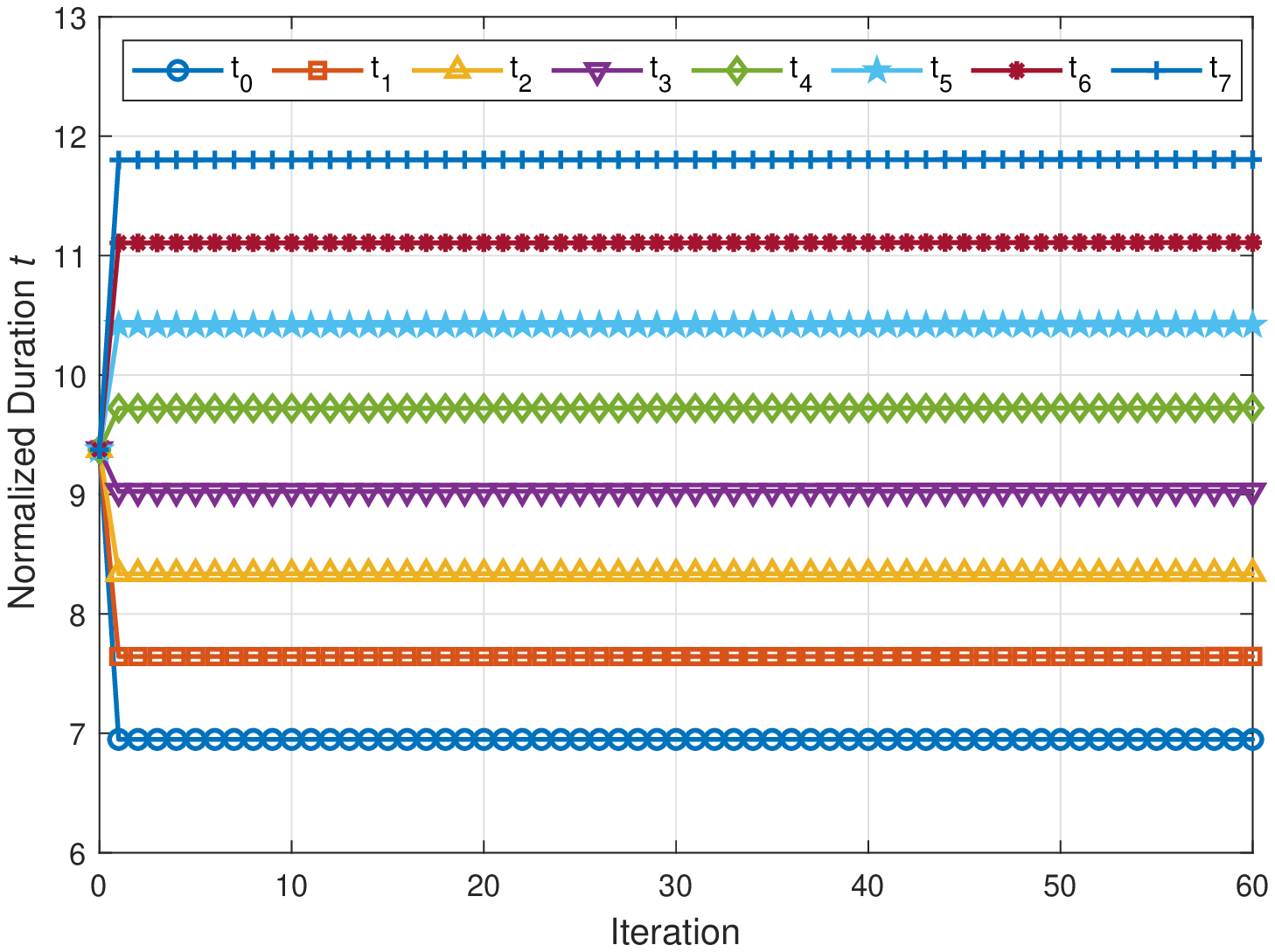}}			
	\end{centering}
	\caption{Convergence of Algorithm~\ref{algo:acs} ($B = 8$ and $\mathsf{E} = 300$): (a) $\vect{i}^*$ for $\vect{i}^{(0)} = (1, \ldots, 1)$, (b) $\vect{t}^*$ for $\vect{i}^{(0)} = (1, \ldots, 1)$, (c) $\vect{i}^*$ for $\vect{i}^{(0)} = (2, \ldots, 2)$, and (d) $\vect{t}^*$ for $\vect{i}^{(0)} = (2, \ldots, 2)$.}
	\label{fig:iter} 
	\end{figure*}

	\begin{figure}[t]
	\centering
	\includegraphics[width=0.45\textwidth]{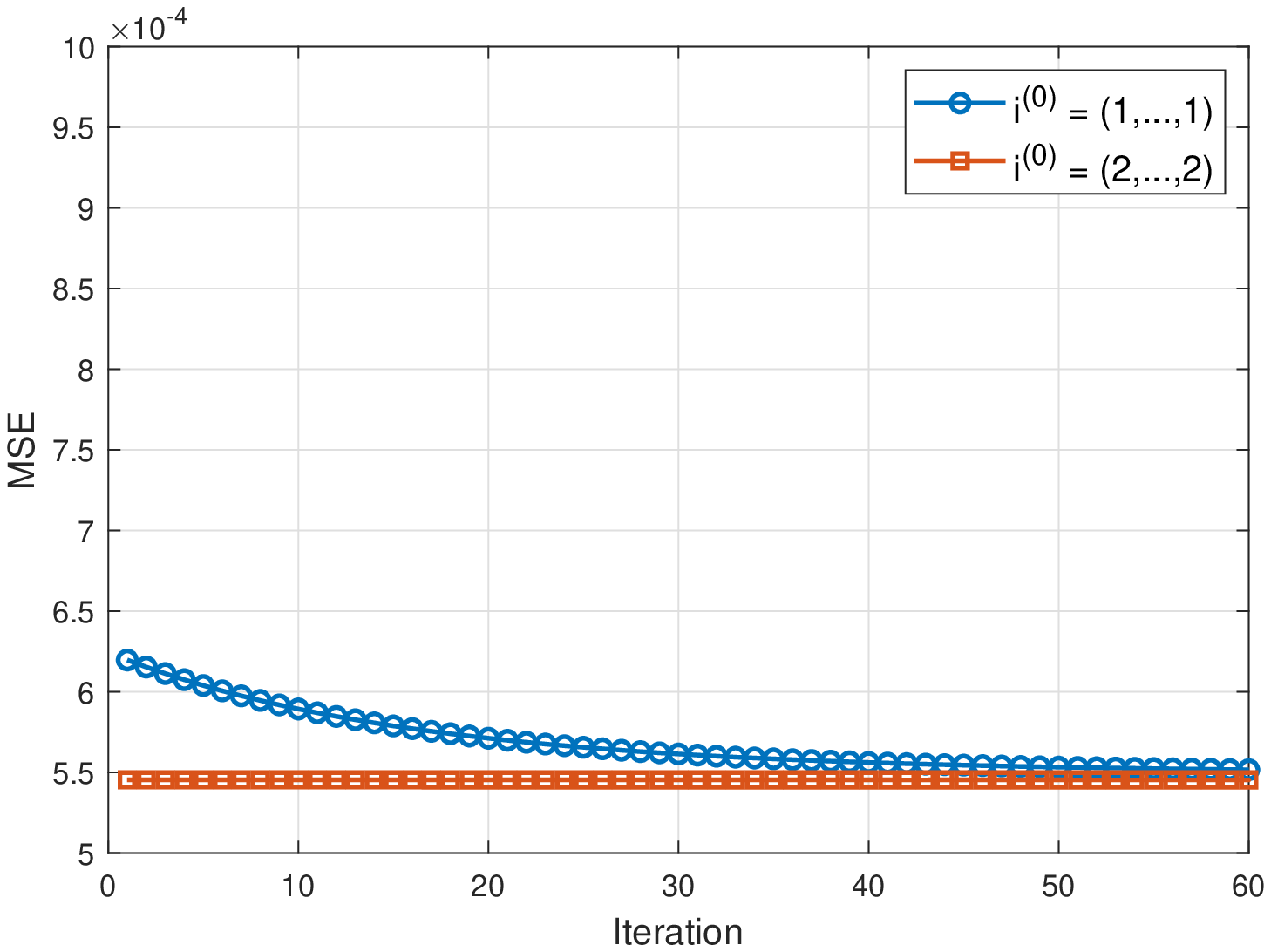}
	\caption{The MSE comparison for $\vect{i}^{(0)} = (1, \ldots, 1)$ and $\vect{i}^{(0)} = (2, \ldots, 2)$ ($B = 8$ and $\mathsf{E} = 300$).}
	\label{fig:iter_mse}
	\end{figure}
	
	Fig.~\ref{fig:iter} characterizes the convergence of Algorithm~\ref{algo:acs}. The convergence speed depends on the starting point $\vect{i}^{(0)}$. For both $\vect{i}^{(0)} = (1,\ldots,1)$ and $\vect{i}^{(0)} = (2,\ldots,2)$, Algorithm~\ref{algo:acs} converges; however, the convergence speed of $\vect{i}^{(0)} = (1,\ldots,1)$ is slower than that of $\vect{i}^{(0)} = (2,\ldots,2)$. As shown in Corollary~\ref{thm:convergence_fast}, the starting point $\vect{i}^{(0)} = (2,\ldots,2)$ guarantees the fastest convergence (see Fig.~\ref{fig:iter}(c) and (d)). Fig.~\ref{fig:iter_mse} compares the MSEs of $\vect{i}^{(0)} = (1, \ldots,1)$ and $\vect{i}^{(0)} = (2,\ldots,2)$. We observe that the MSE for $\vect{i}^{(0)} = (2,\ldots,2)$ is better than that for $\vect{i}^{(0)} = (1,\ldots,1)$. The gap between these two MSE is vanishing as iterations progress.

%
%
	
	\section{Conclusion}\label{sec:conclusion}
	
	We proposed an information-theoretic approach to improving MRAM's write energy efficiency. After formulating the biconvex optimization problem, we proposed the iterative algorithm to solve the biconvex problem, which attempts to minimize the MSE under a refresh power budget. Also, we proved that the proposed algorithm converges and it can reduce the MSE exponentially. The proposed optimization scheme can be extended in future work to coded information representations, where redundancy is added to the written values to further improve the fidelity. 
	

	\appendices
	
	\section{Proof of Lemma~\ref{thm:min_wfp}}\label{pf:min_wfp}
	
	It is clear that $i^*$ and $t^*$ satisfy $i^2 t = \mathcal{E}$ to maximize $(i-1)t$. Then, we can set $t = \frac{\mathcal{E}}{i^2}$ and the corresponding objective function is given by
	\begin{equation}
	g(i) = (i-1)t = \mathcal{E} \cdot \frac{i-1}{i^2}.  
	\end{equation} 
	Since $g'(i) = \mathcal{E} \cdot \frac{2-i}{i^3}$, $g'(2) = 0$ and $g'(i) < 0$ for $i > 2$. Hence, $g(i)$ is maximized when $i^* = 2$ and $t^* = \frac{\mathcal{E}}{4}$. 
		
	\section{Proof of Theorem~\ref{thm:min_mse_t}}\label{pf:min_mse_t}
	
	The corresponding KKT conditions are as follows:
	\begin{align}
	\sum_{b=0}^{B-1}{i_b^2 t_b} &\le \mathcal{E}, \quad \nu \ge 0, \quad
	\nu \cdot \left(\sum_{b=0}^{B-1}{
		i_b^2 t_b} - \mathcal{E}\right) = 0, \label{eq:cr1_KKT_t_1} \\
	t_b &\ge 0, \quad \lambda_b \ge 0, \quad \lambda_b t_b = 0 \label{eq:cr1_KKT_t_2} \\
	\frac{\partial L_1}{\partial t_b} &= -2 \cdot 4^b(i_b-1) e^{-2(i_b - 1)t_b} + \nu i_b^2 - \lambda_b = 0 \label{eq:cr1_KKT_t_3}
	\end{align}
	for $b \in [0, B-1]$. From \eqref{eq:cr1_KKT_t_3}, $\lambda_b$ is given by
	\begin{equation} \label{eq:cr1_KKT_t_lambda}
	\lambda_b = i_b^2 \left(\nu  - \frac{2 \cdot 4^b(i_b-1)}{i_b^2} \cdot e^{-2(i_b - 1)t_b} \right). 
	\end{equation}
	Suppose that $\nu = 0$. Then $\lambda_b < 0$ because of $i_b \ge 1 + \epsilon$, which violates the condition of $\lambda \ge 0$. Hence, $\nu \ne 0$ and 
	\begin{equation} \label{eq:cr1_KKT_energy}
	\sum_{b=0}^{B-1}{i_b^2 t_b} = \mathcal{E}.
	\end{equation}   
		
	From \eqref{eq:cr1_KKT_t_2} and \eqref{eq:cr1_KKT_t_lambda}, 
	\begin{align}
	\lambda_b t_b  = i_b^2 \cdot t_b \left\{\nu  - \frac{2 \cdot 4^b(i_b-1)}{i_b^2} \cdot e^{-2(i_b - 1)t_b} \right\} = 0. \label{eq:cr1_KKT_t_slack_1}
	\end{align}
	Because of $\lambda_b \ge 0$ and $i_b \ge 1 + \epsilon$, we obtain 
	\begin{equation} \label{eq:cr1_KKT_t_nu}
	\nu \ge \frac{2 \cdot 4^b(i_b-1)}{i_b^2} \cdot e^{-2(i_b - 1)t_b}. 
	\end{equation}	
		
	If $\nu \ge \frac{2 \cdot 4^b(i_b-1)}{i_b^2}$, then $t_b = 0$. Otherwise (i.e., $t_b > 0$ and $\lambda_b = 0$), then $\nu = \frac{2 \cdot 4^b(i_b-1)}{i_b^2} \cdot e^{-2(i_b - 1)t_b}$. Because of $e^{-2(i_b - 1)t_b} < 1$ for $t_b > 0$, which contradicts to $\nu \ge \frac{2 \cdot 4^b(i_b-1)}{i_b^2}$. 
	
	If $\nu < \frac{2 \cdot 4^b(i_b-1)}{i_b^2}$, then $t_b = 0$ is not allowed because of \eqref{eq:cr1_KKT_t_nu}. Hence, $t_b > 0$ and $\lambda_b = 0$. By \eqref{eq:cr1_KKT_t_lambda} and $i_b \ge 1+\epsilon$,
	\begin{equation} \label{eq:cr1_KKT_t_nu_eq}	
	\nu = \frac{2 \cdot 4^b(i_b-1)}{i_b^2} \cdot e^{-2(i_b - 1)t_b},  
	\end{equation}
	which results in
	\begin{equation}
	t_b^* = \frac{1}{2(i_b - 1)} \log\left(\frac{1}{\nu} \cdot \frac{2\cdot4^b (i_b - 1)}{i_b^2}\right). 
	\end{equation}	

	\section{Proof of Theorem~\ref{thm:min_mse_i}}\label{pf:min_mse_i}

	The corresponding KKT conditions are as follows:
	\begin{align}
	\sum_{b=0}^{B-1}{i_b^2 t_b} &\le \mathcal{E}, \quad \nu' \ge 0, \quad
	\nu' \cdot \left(\sum_{b=0}^{B-1}{
		i_b^2 t_b} - \mathcal{E}\right) = 0, \label{eq:cr1_KKT_i_1} \\
	i_b &\ge 1 + \epsilon, \quad \lambda_b' \ge 0, \quad \lambda_b' \{i_b - (1 + \epsilon)\}  = 0 \label{eq:cr1_KKT_i_2} \\
	\frac{\partial L_2}{\partial i_b} &= -2 \cdot 4^b t_b e^{-2(i_b - 1)t_b} + 2 \nu' t_b i_b  - \lambda_b' = 0 \label{eq:cr1_KKT_i_3}
	\end{align}
	for $b \in [0, B-1]$. From \eqref{eq:cr1_KKT_i_3}, $\lambda_b'$ is given by
	\begin{equation} \label{eq:cr1_KKT_i_lambda}
	\lambda_b' = 2 t_b i_b \left(\nu' - \frac{4^b e^{- 2t_b (i_b - 1)}}{i_b} \right).
	\end{equation}
	Suppose that $\nu' = 0$. Then, $\lambda_b' = -2t_b \cdot 4^b e^{-2t_b(i_b -1 )} \le 0$, which is true only if $t_b = 0$ for all $b \in [0, B-1]$. Since this is a trivial case, we focus on $\nu' \ne 0$ and $\sum_{b=0}^{B-1}{i_b^2 t_b} = \mathcal{E}$.
	
	If $t_b = 0$, then the corresponding $i_b$ affects neither the MSE nor the energy. Hence, we suppose that $t_b \ne 0$. If $\lambda_b' = 0$, then 
	\begin{equation}\label{eq:cr1_KKT_i_nu}
		\nu' = \frac{4^b e^{-2t_b (i_b - 1)}}{i_b},
	\end{equation} 
	which is equivalent to
	\begin{equation}\label{eq:cr1_KKT_i_lambert}
		\frac{2\cdot 4^b t_b e^{2t_b}}{\nu'} = 2t_b i_b e^{2t_b i_b} = z e^z
	\end{equation}
	where $z = 2t_b i_b$. Then, $z = 2t_b i_b = W\left(\frac{2\cdot 4^b t_b e^{2t_b}}{\nu'}\right)$ where $W(\cdot)$ denotes the Lambert W function~\cite{Corless1996lambert}.  
	Hence, 
	\begin{equation} \label{eq:cr1_KKT_i_opt1}
	i_b^* = \frac{1}{2t_b} W\left(\frac{2\cdot 4^b t_b e^{2t_b}}{\nu'}\right). 
	\end{equation}
	Note that $i_b^* = 1+ \epsilon$ for $\nu' = \frac{4^b e^{-2t_b \epsilon}}{1 + \epsilon}$ because of $W(ze^z) = z$.   
		
	Suppose that $g(i_b) = \frac{4^b e^{-2t_b (i_b - 1)}}{i_b}$ in \eqref{eq:cr1_KKT_i_lambda}. Because of $\frac{\text{d}g(i_b)}{\text{d}i_b} < 0$, $g(i_b)$ is a monotonically decreasing function. If $\nu' > \frac{4^b e^{-2t_b \epsilon}}{1 + \epsilon} = g(1 + \epsilon)$, then $\lambda_b' \ne 0$ and $i_b^* = 1 + \epsilon$ by \eqref{eq:cr1_KKT_i_2}.

	\section{Proof of Corollary~\ref{thm:convergence_fast}}\label{pf:convergence_fast}
	
	If $i_b^{(0)} = 2$ and $t_b^{(1)} > 0$ for all $b \in [0,B-1]$, then we obtain the following solution by solving \eqref{eq:min_mse_t}. 
	\begin{equation}
	t_b^{(1)} = \frac{1}{2}\log{\left(\frac{4^b}{2\nu}\right)},
	\end{equation} 
	which follows from \eqref{eq:min_mse_sol_t}. We will show that $(\vect{i}^{(0)}, \vect{t}^{(1)})$ also satisfies all the KKT conditions for \eqref{eq:min_mse_t} (i.e., \eqref{eq:cr1_KKT_i_1}--\eqref{eq:cr1_KKT_i_3} in Appendix~\ref{pf:min_mse_i}). First, $(\vect{i}^{(0)}, \vect{t}^{(1)})$ satisfies \eqref{eq:cr1_KKT_i_1} which is equivalent to \eqref{eq:cr1_KKT_t_1}. In addition, $i_b^{(0)} = 2$ satisfies \eqref{eq:cr1_KKT_i_2} and makes $\lambda_b' = 0$ for all $b\in[0,B-1]$. Then, \eqref{eq:cr1_KKT_i_3} will be
	\begin{equation} \label{eq:convergence_fast_KKT}
	-2\cdot 4^b t_b^{(1)} e^{-2t_b^{(1)}} + 4 \nu' t_b^{(1)} = 0. 
	\end{equation}
	Suppose that $\vect{i}^{(1)} = \vect{i}^{(0)}$. Then, \eqref{eq:cr1_KKT_i_nu} is modified to $\nu' = \frac{4^b}{2}e^{-2 t_b^{(1)}}$, which satisfies \eqref{eq:convergence_fast_KKT}. Thus, $(\vect{i}^{(0)}, \vect{t}^{(1)})$ satisfies all the KKT conditions of \eqref{eq:min_mse_t} and \eqref{eq:min_mse_i}.

	\section{Proof of Lemma~\ref{thm:positive_t} and Theorem~\ref{thm:mse_improvement}}\label{pf:mse_improvement}
	
	From \eqref{eq:min_mse_sol_t_i0}, we observe that $\widetilde{t}_b > 0$ for all $b \in [0, B-1]$ if $\nu < \frac{1}{2}$. By \eqref{eq:cr1_KKT_energy}, $(\vect{i}^{(0)}, \widetilde{\vect{t}} )$ satisfies
	\begin{equation}
	\sum_{b=0}^{B-1}{4 t_b} = \sum_{b=0}^{B-1}{2 \log{\left(\frac{1}{\nu}\cdot \frac{4^b}{2}\right)}} = \mathcal{E}, 
	\end{equation}
	which results in 
	\begin{equation}\label{eq:pf_nu}
	\nu = 2^{B-2} \cdot \exp\left( -\frac{\mathcal{E}}{2B} \right). 
	\end{equation}
	Then, the condition $\nu < \frac{1}{2}$ is equivalent to
	\begin{equation} \label{eq:pf_energy_condition}
	\mathcal{E} > 2B(B-1) \log{2}. 
	\end{equation}
	Hence, $\widetilde{t}_b > 0$ for all $b \in [0, B-1]$ 
	if \eqref{eq:pf_energy_condition} holds. By \eqref{eq:min_mse_sol_t_i0} and \eqref{eq:pf_nu}, we obtain \eqref{eq:positive_t}. 
	
	By \eqref{eq:mse} and \eqref{eq:positive_t}, 
	\begin{align}
	\mathsf{MSE}\left(\vect{i}^{(0)}, \widetilde{\vect{t}}\right) & = c  \cdot \sum_{b=0}^{B-1}{4^b \exp\left(-2 \widetilde{t}_b\right)} \nonumber \\
	& = c \cdot B \cdot 2^{B-1} \exp\left(- \frac{\mathcal{E}}{2B}\right). \label{eq:pf_mse_nonuniform}
	\end{align}	
	The uniform energy allocation of $(\vect{i}^{(0)}, \vect{t}^{(0)})$ results in 
	\begin{equation}
	\mathsf{MSE}\left(\vect{i}^{(0)}, \vect{t}^{(0)}\right) = c \cdot \frac{4^B - 1}{3} \exp\left(-\frac{\mathcal{E}}{2B}\right). \label{eq:pf_mse_uniform}
	\end{equation}
	From \eqref{eq:pf_mse_nonuniform} and \eqref{eq:pf_mse_uniform}, we obtain \eqref{eq:mse_improvement}.

	
	\section*{Acknowledgment}
	
	The work of Yuval Cassuto was partly supported by the US-Israel Binational Science Foundation, and by the Israel Science Foundation.
	
	
	
	\bibliographystyle{IEEEtran}
	\bibliography{abrv,mybib}

\end{document}